\documentclass[conference]{IEEEtran}
\IEEEoverridecommandlockouts
\usepackage{cite}
\usepackage{amsmath,amssymb,amsfonts}
\usepackage{algorithmic}
\usepackage{graphicx}
\usepackage{textcomp}
\usepackage{xcolor}
\usepackage{hyperref}
\usepackage{ctable}
\usepackage{balance}

\def\BibTeX{{\rm B\kern-.05em{\sc i\kern-.025em b}\kern-.08em
    T\kern-.1667em\lower.7ex\hbox{E}\kern-.125emX}}
\begin{document}

\title{An ELEGANT dataset with Denial of Service and Man in The Middle attacks\\
\thanks{The data collected resulted from the funding of FedFire+ Open calls - large experiments.}
}

\author{\IEEEauthorblockN{Bruno Sousa}
\IEEEauthorblockA{\textit{CISUC, DEI} \\
\textit{University of Coimbra}\\
Coimbra, Portugal \\
bmsousa@dei.uc.pt}
\and
\IEEEauthorblockN{Tiago Cruz}
\IEEEauthorblockA{\textit{CISUC, DEI} \\
\textit{University of Coimbra}\\
Coimbra, Portugal \\
tjcruz@dei.uc.pt}
\and
\IEEEauthorblockN{Miguel Arieiro}
\IEEEauthorblockA{\textit{CISUC, DEI} \\
\textit{University of Coimbra}\\
Coimbra, Portugal \\
marieiro@student.dei.uc.pt}
\and
\IEEEauthorblockN{Vasco Pereira}
\IEEEauthorblockA{\textit{CISUC, DEI} \\
\textit{University of Coimbra}\\
Coimbra, Portugal \\
vasco@dei.uc.pt}
}

\maketitle

\begin{abstract}
This document describes a dataset with diverse types of Denial of Service (DoS) attacks and Man-in-the-Middle (MiTM) attacks. The data is available online and reachable via the DOI  \href{https://dx.doi.org/10.21227/mewp-g646}{10.21227/mewp-g646}. This document describes the data collection process and provides useful information on how such data can be employed to devise models for cybersecurity in critical infrastructures using Programmable Logic Controllers (PLCs)
\end{abstract}


\section{Introduction}
Modern automation technologies have become pervasive~\cite{ref_4}, playing a crucial part in ensuring the delivery and availability of several essential services. As a result, operators and service utilities are often compelled by stakeholders, governmental bodies, as well as regulatory, standardisation and steering organisations, to improve service quality, security and reliability. In this regard, the constant monitoring of control elements (i.e., Programmable Logic Controllers - PLCs) is crucial for management and also to detect anomalous behaviour in Industrial Automation Control Systems (IACS). 

PLCs, running protocols like Modbus/TCP~\cite{modbus_tcp}, are often targeted by diverse types of attacks. These may include amplification or network flooding attempts, for Denial of Service (DoS)~\cite{DDoSsurveySmartGrid} purposes, as well as more sophisticated techniques, such as Man in the Middle (MiTM) attacks, using techniques like Address Resolution Protocol (ARP) poisoning to tamper with process information (e.g. water pumps, water levels, temperature sensors, turbine speed sensors, etc) and/or cause loss of visibility. As an example of the latter case, an attacker might leverage MiTM techniques to manipulate temperature sensor measurements: while a sensor may report a real value of 120ºC, an attacker may be able to modify inflight telemetry to deceive PLCs and other Human Machine Interface (HMI) elements to assume and report a temperature reading of 40ºC.

This paper documents one of the outcomes achieved in the ELEGANT project, funded by the 7\textsuperscript{th} Open call of Fed4Fire+ for large experiments\footnote{\href{https://www.fed4fire.eu/event/7th-fed4fire-open-call-large-experiments/}{7th OpenCall Fed4Fire+ Large Experiments.}}, namely the dataset with DDoS and MiTM attacks that is publicly available at the IEEE DataPort repository (DOI:~\href{https://dx.doi.org/10.21227/mewp-g646}{10.21227/mewp-g646}). The dataset includes flooding and amplification DDoS attacks performed from single and multiple nodes, as well as ARP-poisoning based MiTM attacks.

\section{Collection Process}
\label{sec:collectionProcess}

\subsection{Overall architecture}
\begin{figure}[!htbp]
  \centering
	\includegraphics[width=0.9\columnwidth]{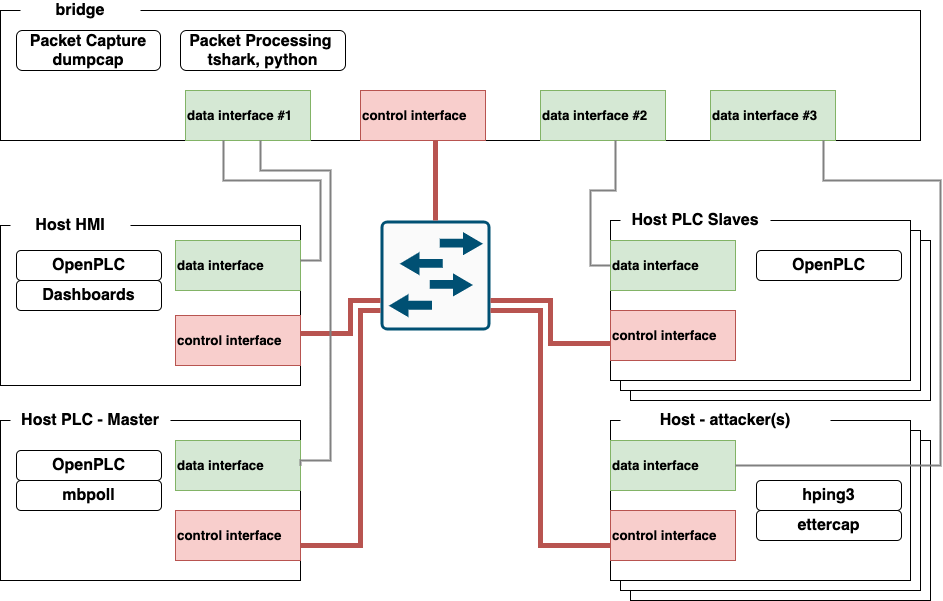}
	\caption{Overall architecture with PLCs}
	\label{fig:arch}
\end{figure}

The base architecture that was deployed in Fed4Fire+ testbeds (virtual Wall2 and Grid5000) and used to collect the data available in the dataset is pictured in Figure~\ref{fig:arch}.

All the components have a data and a control interface. All the relevant information, namely the Modbus TCP~\cite{modbus_tcp} data packets exchanged in the scope of the process control tasks, were collected in the data interfaces. The capture was performed in the bridge node using the dumpcap tool~\cite{wireshark}.

The PLC master node queries other PLC nodes regarding the information of sensors stored in specific registers. PLC slaves have sensors information  which is stored in the holding registers as per the PLC internal mechanism. It should be noticed that multiple PLCs exist in the testbed. 
All the PLC are based on the OPenPLC v3 version~\cite{OpenPLC}.

\subsection{PLC Modbus settings}
\label{sec:collectionProcess:PLC}

Regarding the reference topology, there are both horizontal (PLC-PLC) and vertical (PLC-HMI) communication parterns. PLCA1 has multiple configured slaves, which are polled in 100ms intervals. Such polling can be reduced in real-world deployments, for instance to values around 40ms, leading to rates around 25pkts/s~\cite{Niedermaier2019}. To determine the rates in the attacks, we consider 25pkts/s as the reference rate for the Modbus TCP polling process.

\begin{figure}[!htbp]
  \centering
	\includegraphics[width=0.9\columnwidth]{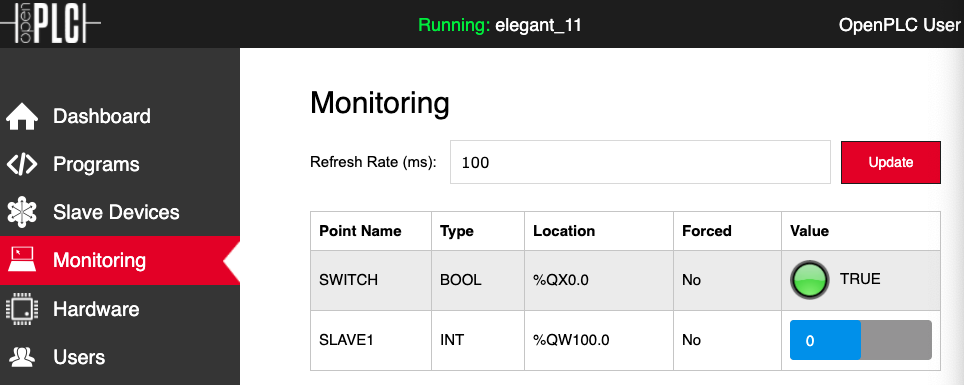}
	\caption{ELEGANT Water level program variables}
	\label{fig:ELEGANTv11pgm}
\end{figure}

PLC master queries the diverse PLCs (e.g. PLC slave) which implement the logic for querying information from sensors, also performing actuation tasks based on such information. Figure~\ref{fig:ELEGANTv11pgm} illustrates the functionality of the program running at the PLCs. The underlying logic considers a process for water level control in a tank, using a level sensor and a water pump (labelled as SWITCH in Figure.~\ref{fig:ELEGANTv11pgm}) which is activated if a certain threshold is achieved. 

\subsection{Denial of Service attacks}

The Deninal of Service (DoS) attacks have been performed considering multiple configurations. Two traffic rate classes were considered for the inter-arrival packet times. Rate 1 includes packet interval rates in a ratio of 10 times bigger the normal rate of the Modbus TCP polling process ($10 x normalRate$). Rate 2 assumes the maximum flood that can be performed with the attack node (\textit{-\--flood} option), as summarized in Table~\ref{table:ddos_rate_attacks}. 

\ctable[
cap=DoS attack rate settings ,
caption={DoS attack rate settings},
label=table:ddos_rate_attacks,
pos=!htb,
doinside=\footnotesize,
center,
nostar
]{m{1cm} m{2cm} m{3.5cm}  }{
}{ \FL
\textbf{Rate} & \textbf{Interval (ms)} & \textbf{N. packets}   \ML 
 1 &  400  & 2500 \\
 2 &  $<1$ & Max. supported by node
 \LL
}

Table~\ref{table:ddos_attacks} summarizes the tests that were performed, illustrating the time periods (CET timezone). Each test was performed using the hping3 tool~\cite{hping3} with the ability to randomize source node IP (\textit{--rand-source} option). The DoS targeted the PLCA1, which acts as a master, polling information from other devices (other PLCs, such as PLCA2).

The DoS attack types encompass flooding attempts, which target the modbus TCP port (502) with a packet size of 120 bytes (size considered as per the size of the  traffic in the polling process of the program running in the PLCs, recall sub-section~\ref{sec:collectionProcess:PLC}). DoS-based amplification attacks rely on UDP traffic for the same port, but with a size of 60 bytes.

\ctable[
cap=DoS attacks,
caption={DoS attacks},
label=table:ddos_attacks,
pos=!htb,
doinside=\footnotesize,
center,
star
]{m{2cm} m{2cm} m{1cm} m{2cm} m{3cm} m{5cm}  }{
}{ \FL
\textbf{Node (S)ingle, (M)ultiple} & \textbf{type} & \textbf{rate} & \textbf{Date} & \textbf{Hour interval (CET)} & \textbf{Files/ Timestamp}   \ML 
 S &  Flood & 1 & 2021/03/13  & 19:44-19:47 & 00001-00008 \\
 S &  Flood & 1 & 2021/03/13  & 19:55-19:58 & 00008-00016 \\
 
S &  Flood & 2 & 2021/03/13  & 20:02-20:05 & 00016-00249 \\
S &  Flood & 2 & 2021/03/13  & 20:10-20:13 & 00249-00485 \\

S &  Amp & - & 2021/03/13  & 20:20-20:24 & 00485-00661 \\
S &  Amp & - & 2021/03/13  & 20:29-20:32 & 00661-00841 \\
\hline
M &  Flood & 1 & 2021/03/14  & 01:02-01:05 & 20210314005755- 20210314010515 \\
M &  Flood & 1 & 2021/03/14  & 01:37-01:40 & 20210314013327- 20210314014020 \\
M &  Flood & 2 & 2021/03/14  & 01:12-01:15 & 20210314010944- 20210314011526 \\
M &  Flood & 2 & 2021/03/14  & 01:19-01:22 & 20210314011526- 20210314012226 \\
M &  Amp & - & 2021/03/14  & 01:44-01:47 & 20210314014020- 20210314014821 \\
M &  Amp & - & 2021/03/14  & 01:51-01:55 & 20210314014821- 20210314015502
\LL
}

The beginning and end of the attack timeframes are also documented on the respective pcap files, collected with the dumpcap tool, which was configured to dump network traffic in a ring buffer with file sizes between 10 and 20MB.

\begin{figure}[!htbp]
  \centering
	\includegraphics[width=0.9\columnwidth]{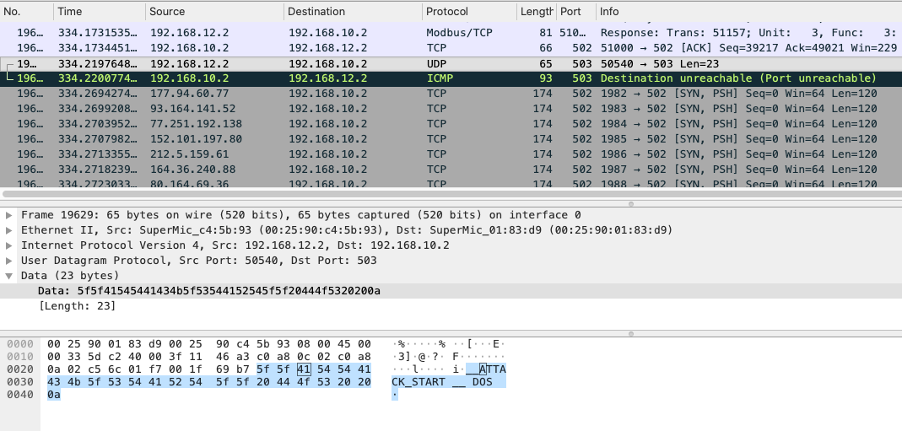}
	\caption{Start label of the attack in dataset}
	\label{fig:startDdos}
\end{figure}

The start of the attack is present in the pcap files, as illustrated in Figure~\ref{fig:startDdos}. In particular, a UDP data packet with a length of 23 bytes is sent to to port 503, on the destination PLC. The data contained if the data field of the packet contains the string "ATTACK\_\_START\_\_DDOS".

\begin{figure}[!htbp]
  \centering
	\includegraphics[width=0.9\columnwidth]{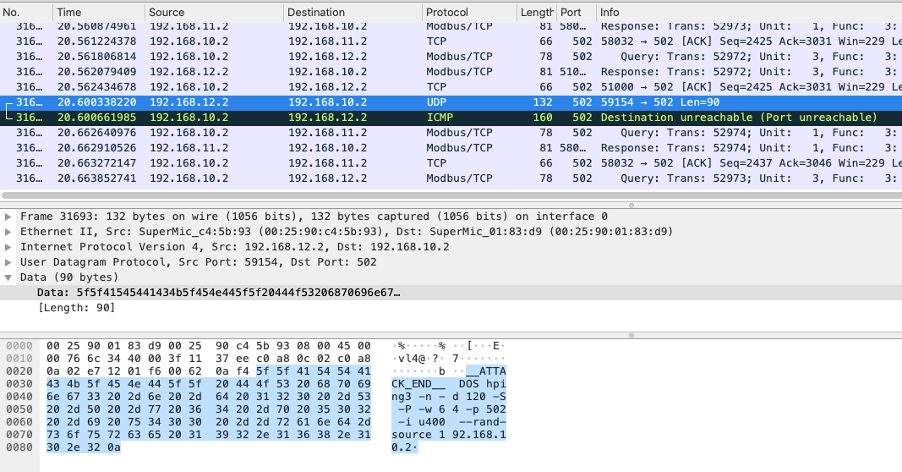}
	\caption{End label of the attack in dataset}
	\label{fig:endDdos}
\end{figure}

The end of the attack, as illustrated in Figure~\ref{fig:endDdos}, consists of a UDP message sent to port 502, with a size of around 90 bytes. The data field contains information of the attack that was performed. For instance, as illustrated in Figure~\ref{fig:endDdos}, one can observe the label "ATTACK\_\_END\_\_ DOS hping3". 

\subsection{Man in the middle attacks}

The technique selected to deploy the MiTM attacks executed within the scope of the ELEGANT dataset extraction process is based on the ARP poisoning technique.  ARP is a stateless protocol, designed in such a way what it implicitly relies on unprotected Layer 2/3 mechanisms to provide a dynamic association mechanism that binds MAC to IP addresses within the same link-layer domain.

By design, ARP will cache any replies, whether solicited or not, overwriting non-expired entries when new replies are received. Due to the lack of authentication and authorisation mechanisms, ARP constitutes a weak spot that can be exploited to implement MiTM attacks~\cite{Foglietta2019} against IPv4 network environments.

ARP-poisoning MiTM attacks can be deployed for different purposes~\cite{Rosa2017}. For instance, they may provide the means for an attacker to discreetly scout a target infrastructure, capturing network traffic for analysis as part of the early planning or attack preparation stages. But these attacks can also be useful in offensive roles, allowing attackers to hijack Modbus TCP connections and change/mask relevant information, effectively blinding the operator and disrupting process operation.

\begin{figure}[!htbp]
  \centering
	\includegraphics[width=0.9\columnwidth]{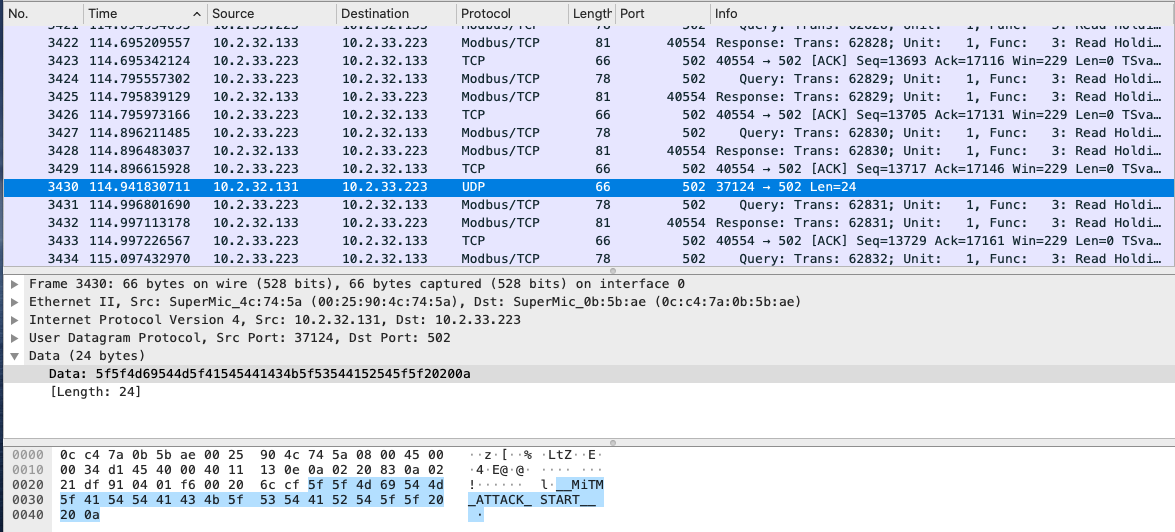}
	\caption{Start label of the MiTM attack in dataset}
	\label{fig:startMiTM}
\end{figure}

The Man in the Middle attacks have been performed in two flavors, as summarized in Table~\ref{table:MiTM_attacks}, one considering only the A-ARP poisoning, and another considering the F-full chain of the attack, which modifies data regarding Modbus/TCP protocol. The attacks have been performed to last around 10 minutes, with the exception of the test in the full chain in run 2, which lasts around 6 minutes. A long period with data without MiTM attacks is also provided (labelled as no attack in Table~\ref{table:MiTM_attacks}).
The attacks are identified in the traces with the protocol UDP at destination port 502 and with the string \"\_\_MiTM\_ATTACK\_ START\_\_" in the data field, as illustrated in Figure.~\ref{fig:startMiTM}. 

\begin{figure}[!htbp]
  \centering
	\includegraphics[width=0.9\columnwidth]{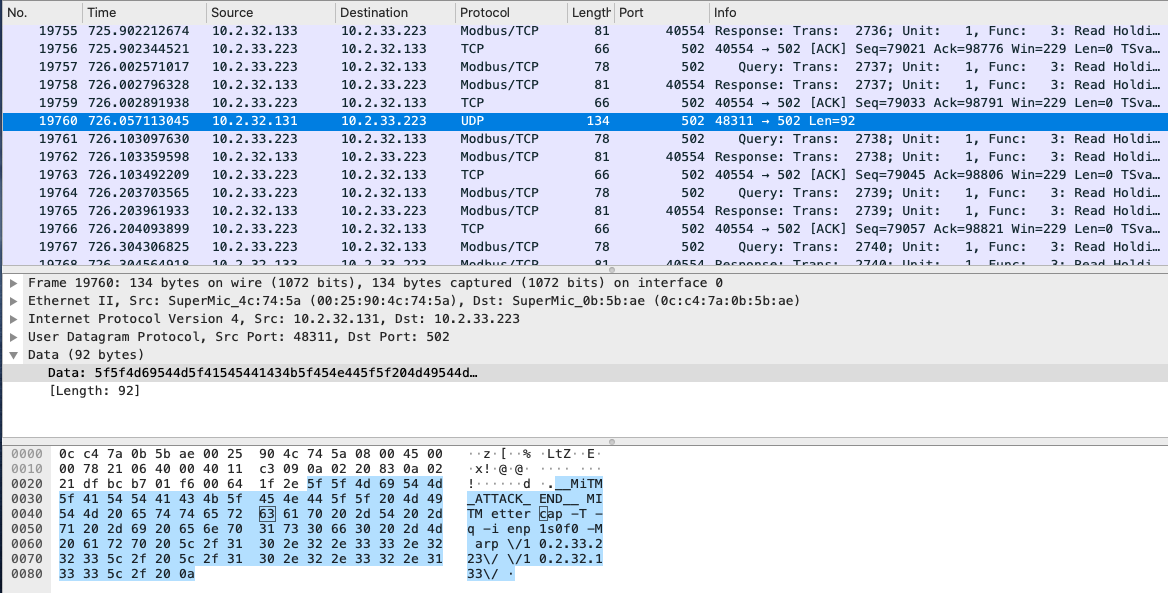}
	\caption{End label of the MiTM attack in dataset}
	\label{fig:endMiTM}
\end{figure}

The end of attacks are identified with packets sent to destination port 502 and with the data field with a string regarding the pattern \"\_\_MiTM\_ATTACK\_ END\_\_".

\ctable[
cap=MiTM attacks,
caption={MiTM with (A)RP poisoning and (F)ull Chain attacks},
label=table:MiTM_attacks,
pos=!htb,
doinside=\footnotesize,
center,
nostar
]{m{1cm} m{0.3cm} m{1cm} m{1.5cm} m{2.5cm}  }{
}{ \FL
\textbf{type (A),(F) } & \textbf{run} & \textbf{Date} & \textbf{Hour interval (CET)} & \textbf{Files/ Timestamp}   \ML 
A & 1 & 2021/03/18   & 03:40-03:50 & 20210318034103 \\
A & 2 & 2021/03/18 & 08:50-09:02 & 20210318084952 \\
\hline
F & 1 & 2021/03/18 & 09:52-10:02 & 20210318095210 \\
F & 2 & 2021/03/18 & 10:04-10:10 & 20210318100414 \\
F & 3 & 2021/03/18 & 10:16-10:26 & 20210318101444 \\
\hline
\multicolumn{2}{c}{no attack} & 2021/03/18 & 03:53-08:46 & 20210318035345- 20210318080920
\LL
}

The MiTM attack is performed with the ettercap tool~\cite{ettercap} in text mode with the option -M arp. The full chain of the attack with packet manipulation is enabled with a filter (option -F name\_filter) to manipulate the information in the Modbus reply messages regarding the values of the records.

\section{Dataset structure and files}

The files available in the dataset are encoded using the pcap format, which stores information in a structured way. The structure considers a global header~\cite{LibPCAPFormat} with the following structure:

\begin{verbatim}
typedef struct pcap_hdr_s {
  guint32 magic_number; /* magic number */
  guint16 version_major; /* major ver. num. */
  guint16 version_minor; /* minor ver. num. */
  gint32  thiszone; /* GMT to local */
  guint32 sigfigs; /* timestamp accuracy */
  guint32 snaplen; /* max length  */
  guint32 network; /* data link type */
} pcap_hdr_t;
\end{verbatim}

More information regarding the meaning of the diverse fields is available at~\cite{LibPCAPFormat}.

\begin{figure}[!htbp]
  \centering
	\includegraphics[width=0.9\columnwidth]{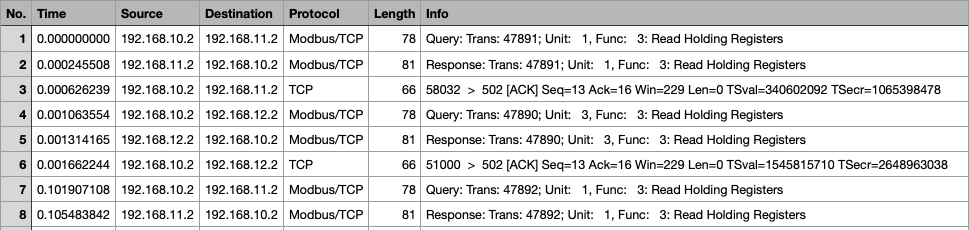}
	\caption{Example of collected data and available in the CSV file}
	\label{fig:figCSVformat}
\end{figure}

In addition, each record (capture packet) is identified by an header, with the following information:
\begin{verbatim}
typedef struct pcaprec_hdr_s {
  guint32 ts_sec; /* timestamp seconds */
  guint32 ts_usec; /* timestamp microsec.*/
  guint32 incl_len; /* number of octets */
  guint32 orig_len; /* length of packet */
} pcaprec_hdr_t;
\end{verbatim}

A CSV file is also available to illustrate the structure of the collected information, see Figure~\ref{fig:figCSVformat}.

The dataset contains several files regarding the DoS attacks, which are organized as follows:
\begin{enumerate}
    \item \textbf{\textit{cap\_001\_2021\_DoS\_example.csv}}- Sample CV file with data from DoS attack in single more and date 1.
    \item \textbf{\textit{single\_flood\_rate\_1\_t1.zip}}- DoS flood attack in single node with rate 1 and run 1.
    \item \textbf{\textit{single\_flood\_rate\_1\_t2.zip}}- DoS flood attack in single node with rate 1 and run 2.
    \item \textbf{\textit{single\_flood\_rate\_2\_t1.zip}}- DoS flood attack in single node with rate 2 and run 1.
    \item \textbf{\textit{single\_flood\_rate\_2\_t2.zip}}- DoS flood attack in single node with rate 2 and run 2.
    \item \textbf{\textit{single\_amp\_t1.zip}}- DoS amplification attack in single node and run 1.
    \item \textbf{\textit{single\_amp\_t2.zip}}- DoS amplification attack in single node and run 2.
    \item \textbf{\textit{multiple\_flood\_0100\_0125.tar.gz}}- DoS flood attack in multiple nodes with rate 1 with run 1, and rate 2 with runs 1 and 2.
    \item \textbf{\textit{multiple\_amp\_0130\_0200.tar.gz}}- DoS flood attack in multiple nodes with rate 1 with run 2, and DoS amplification attack in multiple nodes with run 1 and 2.
\end{enumerate}

The dataset contains several files regarding the DoS attacks, which are organized as follows:
\begin{enumerate}
    \item \textbf{\textit{MiTM\_ARP\_Poisoning.tar.gz}}- MiTM attack with ARP poisoning with run 1 and run 2.
    \item \textbf{\textit{MiTM\_Full\_Chain.tar.gz}}- MiTM attack with Full Chain, including ARP poisoning attack and data manipulation in run 1 and run 2.
    \item \textbf{\textit{normal\_PLC\_traffic.tar.gz}}- Regular ModBus/TCP traffic without attacks.
\end{enumerate}

\section{Conclusion}
We hope this dataset may help others working in the research and development of secure solutions for Critical Infrastructures.

Also, the information herein contained will be updated with more details and developments regarding the achievements and results of the ELEGANT project.

\balance

\bibliographystyle{./bibliography/IEEEtran}
\bibliography{ELEGANTdataRepo.bib}
\end{document}